\documentclass[aps,prb,reprint,superscriptaddress,floatfix]{revtex4-2}

\usepackage[ngerman,english]{babel}

\usepackage{braket}
\usepackage{amsmath}

\usepackage{graphicx}
\usepackage{floatrow}

\usepackage[unicode]{hyperref}
\usepackage{changes}

\begin{document}

\title{Dynamical breaking  of the electron-hole symmetry  in non-equilibrium chiral quantum channels}

\author{Felix Puster}
\email{felix.puster@uni-leipzig.de}
\affiliation{Institut f\"ur Theoretische Physik, Universit\"at Leipzig, Br\"uderstrasse 16, 04103 Leipzig, Germany}

\author{Stefan G. Fischer}
\affiliation{Institut f\"ur Theoretische Physik, Universit\"at Leipzig, Br\"uderstrasse 16, 04103 Leipzig, Germany}

\author{Bernd Rosenow}
\affiliation{Institut f\"ur Theoretische Physik, Universit\"at Leipzig, Br\"uderstrasse 16, 04103 Leipzig, Germany}

\date{\today}

\begin{abstract}

We investigate the relaxation dynamics in a chiral one-dimensional quantum channel with finite-range interactions, driven out of equilibrium by the injection of high-energy electrons. While the distribution of high-energy electrons, after dissipation of some of their energy,   has been examined previously (Ref.~\cite{Fischer.2021}), we study the distribution of charge carriers excited from the channel's Fermi sea during this process. Utilizing a detector to measure the energetic imprint in the Fermi sea downstream of the injection point, we discover an initial symmetry in the distribution of excited electrons and holes relative to the Fermi level. However, this symmetry breaks down with stronger interactions and increased propagation distances, attributed to terms of order four and beyond in the interaction. We provide an intuitive interpretation of these results in terms of interference between states with different numbers of plasmons in the Fermi sea. 
  
\end{abstract}

\maketitle

\begin{figure}[b]
	\includegraphics[width=\textwidth]{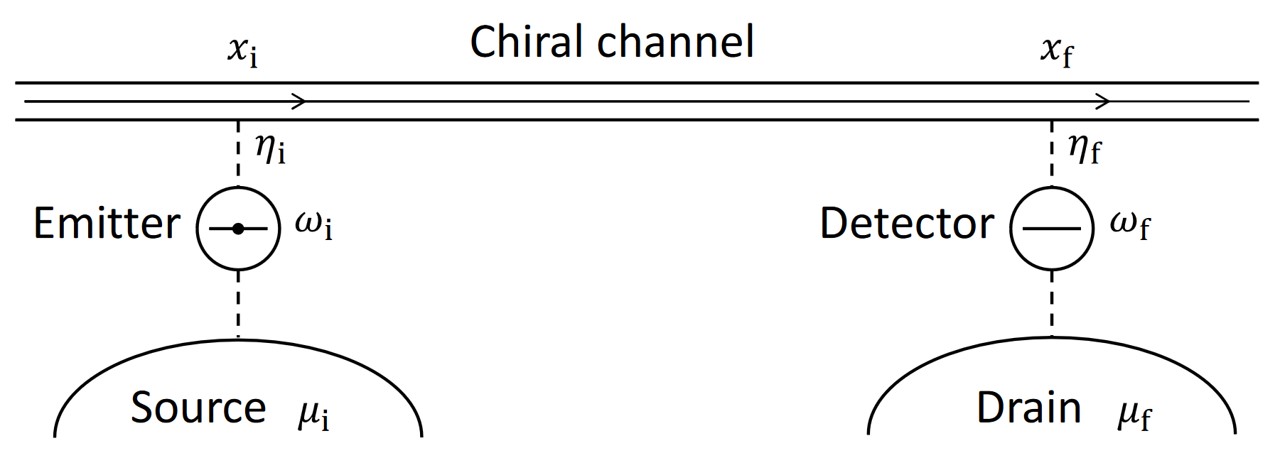}
	\caption{\label{fig:setup}
	Setup: electrons with energy $\omega_{\text{i}}$ are injected from an emitter quantum dot into a chiral quantum channel, driving it out of equilibrium. The chemical potential of the source, $\mu_\mathrm{i}$, is set to ensure constant filling of the emitter quantum dot. The relaxation process of these injected electrons gives rise to excitations within the Fermi sea. After a specified propagation distance, these excitations are detected by a detector quantum dot. To ensure that only excitations are detected, the drain's chemical potential, $\mu_\mathrm{f}$, is chosen equal to the channel's chemical potential, $\mu$. The energy of the detector quantum dot, $\omega_{\text{f}}$, is then varied to determine the density of the excitations. }
\end{figure}

\section{Introduction}

The complex relaxation dynamics in integrable systems offer an array of fascinating features that are not exclusively limited to thermal equilibration~\cite{Polkovnikov.2011,Gogolin.2016,Milletari.2013,Schneider.2017}. Certain experimental settings, such as trapped cold atoms prepared in specific non-equilibrium states, have demonstrated a  lack of intrinsic relaxation~\cite{Kinoshita.2006,Hofferberth.2007}. By utilizing quantum wires, one can explore the relaxation dynamics inherent to the integrable Luttinger liquid model and beyond~\cite{Lerner.2008,Calzona.2017,Calzona.2018b,Strkalj.2019,Deshpande.2010,Barak.2010,Idrisov.2019}. Additionally, quantum Hall edges, hosting chiral one-dimensional quantum channels, present us with further experimental opportunities for creating non-equilibrium states~\cite{Krahenmann.2019,Rodriguez.2020,Kamata.2010,LeSueur.2010,Altimiras.2010}.

In the context of quantum Hall edges that contain multiple channels, considering contact interactions between electrons within these channels provides a sufficient description of relaxation phenomena~\cite{Ferraro.2014,Acciai.2017,Cabart.2018,Rodriguez.2020,LeSueur.2010,Degiovanni.2010,Lunde.2010,Kovrizhin.2011,Levkivskyi.2012,Milletari.2013,Slobodeniuk.2016,Schneider.2017,Duprez.2019}. The resulting dynamics yield nearly-thermal metastable states, recently demonstrated in experimental setups~\cite{Inoue.2014,Itoh.2018}. These metastable states eventually thermalize on longer timescales, when non-linear terms in the dispersion relation gain in importance and break integrability in the system~\cite{Barak.2010,Imambekov.2009,Imambekov.2012,Karzig.2010}.

For a comprehensive understanding of relaxation in single quantum Hall edge channels,  it is necessary to consider finite-ranged interactions~\cite{Degiovanni.2009,Cabart.2018,Kovrizhin.2011,Chalker.2007,Neuenhahn.2008,Neuenhahn.2009}. When the injection energy is below a specific threshold, dictated by the screening length and the interaction-renormalized Fermi velocity (the threshold becomes infinite for contact interactions),  Pauli blockade almost completely suppresses relaxation in these channels~\cite{Cabart.2018,Neuenhahn.2008,Karzig.2012,Fischer.2019}. Above this energy threshold, the relaxation in single channels has recently been explored by considering electron injection at a defined high energy~\cite{Fischer.2021}.  After an initial rapid decay, an unexpected cessation of energy relaxation was found, potentially attributable to an energy mismatch between high-energy electrons and low-energy plasmons. This ultimately results in a metastable state that deviates significantly from equilibrium.

In this study, we focus on the investigation of excitations within the Fermi sea, which occur as a result of the relaxation of high-energy electrons in a single quantum Hall edge channel. To detect these excitations, we employ the resonant level of a quantum dot. The dot operates as a spectrometer for both electrons (above the Fermi level of the channel) and holes (below the Fermi level of the channel). These excitations can tunnel into a drain channel, as depicted in Fig.~\ref{fig:setup}, thereby generating either a positive or negative detector current. The magnitude of this current corresponds to the density of the excitations, enabling us to quantify the perturbations within the channel, which are caused by the energy loss from the injected electron. An intriguing observation from our study is the development of an asymmetry in the densities of excited electrons and holes in the Fermi sea of the channel. This lack of symmetry is presumably induced by the presence of the high-energy electron that accompanies the plasmons excited in the channel. Within a simple model, we interpret the asymmetry between particle and hole excitations in terms of interference between states of the Fermi sea with different plasmon number.

\section{Model and Hamiltonian}

High-energy electrons are injected into the chiral channel through an initial quantum dot, possessing a resonant level at energy $\omega_{\text{i}}$ (see Fig.\ref{fig:setup}). In a scenario with a high injection energy ($\omega_{\text{i}} \gg v/\lambda$), the detector signal resulting from the injected electrons that have dissipated some of their energy can be distinguished from the signal originating from charge carriers that are excited from the Fermi sea~\cite{Fischer.2021}. A subsequent quantum dot detects this signal at an energy level  $\omega_{\text{f}}$. While the signal from injected electrons at $\omega_{\text{f}} \approx \omega_{\text{i}}$ has been explored thoroughly in Ref~\cite{Fischer.2021}, in this study, we examine the signal from charge carriers at $\omega_{\text{f}} \approx \mu$, which have been excited from the Fermi sea during the above-mentioned process.

In this analysis, we will limit our consideration to the case of small tunneling amplitudes ($\eta_{\text{i}/\text{f}}$). Utilizing the Keldysh formalism, we can treat the tunneling Hamiltonians  
\begin{equation}\label{eq:H_tunnel}
H_{t;\text{i}/\text{f}}=\eta_{\text{i}/\text{f}}\int dk e^{-ikx_{\text{i}/\text{f}}} c_k^\dag d_{\text{i}/\text{f}}+e^{ikx_{\text{i}/\text{f}}}d^\dag_{\text{i}/\text{f}}c_k ,
\end{equation}
in a perturbative manner up to the lowest non-vanishing order~\cite{Kane.2003,Takei.2010,Han.2016}. The quantum dots are described by the Hamiltonian  $H_\text{dot}=\omega_{\text{i}} d^\dag_\text{i}d_\text{i}+\omega_{\text{f}}d^\dag_\text{f} d_\text{f}$, and the equilibrium electron system within the channel is characterized by the Hamiltonian
\begin{equation}
	\label{eq:H}
	H = \int dk \,  v k \hat{c}_k^{\dagger}  \hat{c}^{\vphantom{\dagger}}_k +  \frac{1}{4\pi} \int dk dk^\prime dq \, \nu_q^{\vphantom{\dagger}} \hat{c}^\dagger_{k-q} \hat{c}^\dagger_{k'+q} \hat{c}_{k'}^{\vphantom{\dagger}} \hat{c}_{k}^{\vphantom{\dagger}} \ .
\end{equation}
Here, $\nu_q$ represents the Fourier transform of the real space interactions between electrons. These interactions between the injected electrons and those already present in the channel lead to the excitation of plasmons within the Fermi sea of the channel.

\section{Formal expression for the excess electron distribution}

We start by bosonizing the Hamiltonian $H$  in Eq.~(\ref{eq:H}). This process provides us with formulas for the greater ($+$) and lesser ($-$) Green functions of the channel, which can be expressed as $G^{\pm}\left(x, t\right) = G_{0}^{\pm}\left(x, t\right) \exp\left[ S^{\pm}(x, t) \right]$. In this equation, $G_0^{\pm}\left(x, t\right) = 1/2\pi(x - vt \pm i\epsilon)$ represents the non-interacting component which is separated for convenience~\cite{Neuenhahn.2008,Fischer.2021}. The influence of interactions is encapsulated in the exponent. Assuming a zero temperature limit, this component can be defined by the following formula:
\begin{align}
\label{eq:S}
S^{\pm}(x,t) = \int_{0}^{\infty} \frac{dq}{q} \left[ e^{\mp i \left( \omega_q t - qx \right)} - e^{\mp i \left( v q t - qx \right)} \right],
\end{align}
where $\omega_q = vq \left( 1 + \nu_q/2\pi v \right)$ denotes the plasmon dispersion relation.

In the following, we examine interactions characterized by an exponential decay in momentum space, with screening length $\lambda$ and strength $\nu$~\cite{Fischer.2021}, denoted by
\begin{align}
\label{eq:nuq}
\nu^{\text{(exp)}}_{q} = \nu\exp\left(- \lambda |q| \right).
\end{align}
This representation allows us to approximate the corresponding Green functions, thereby facilitating an analytical treatment and improving the efficiency of numerical evaluation.

 Upon employing  Eq.~(\ref{eq:nuq}), the exponent in Eq.~(\ref{eq:S}) can be expressed as a sum of powers of simple poles in the complex $t$ plane~\cite{Fischer.2021},
\begin{align}
\label{eq:Sexplicit}
S^{\text{(exp)}\pm}(x,t) = \sum_{n=1}^{\infty} \frac{1}{n} \left[ \frac{\frac{\nu}{2\pi} t}{x -vt\pm i\lambda n} \right]^n.
\end{align}

We also explore a model  Green function~\cite{Neuenhahn.2008,Fischer.2021} which only contains two velocities instead of a continuum of plasmon velocities
\begin{align}
\label{eq:Gvbarv}
	G^{\pm}_{\rm 2 v} \left(x, t\right) 
	= \frac{1}{2 \pi} \frac{1}{x - vt \pm i\epsilon} \frac{x - vt \pm i\lambda_{\text{c}}}{x - \bar vt \pm i\lambda_{\text{c}}} \ ,
\end{align}
where one pole is determined by the unrenormalized electronic velocity $v$ and another by the interaction renormalized velocity $\bar v = v + \nu/2\pi$, which coincides with the plasmon velocity defined by the derivative of $\omega_q$ at $q=0$.

In both models, we enforce strict chirality, setting the advanced Green function $G^a(x,t)=\Theta(-t)[G^<(x,t)-G^>(x,t)]$ to zero for $x>0$. Assuming small tunneling amplitudes $\eta_{\text{i}/\text{f}}$, the general formula for the distribution of electrons and holes in relation to the channel's ground state~\cite{Kane.2003,Takei.2010,Han.2016}   (which is proportional to the detector signal both above and below the Fermi level)  can be expressed as follows:
\begin{align}
	\label{eq:W}
	&p(x,\omega_{\text{i}},\omega_{\text{f}}) = \nonumber \\ 
	&\quad \frac{ v^2}{2\pi} \int_{-\infty}^{+\infty} \! dt_0  \int_{-\infty}^{+\infty} \! dt_1 \int_{-\infty}^{+\infty} \! dt_2 \,\, e^{i \omega_{\text{f}} t_0} e^{-i \omega_{\text{i}} \left( t_1 - t_2 \right)} \nonumber \\
	&\times G^{-}\left(0, t_1-t_2\right) G^{\alpha}\left(0, -t_0\right) \Big[ \Pi^{-+}(x,0,t_0,t_1,t_2) \nonumber \\ 
	&\qquad \qquad \qquad \qquad \qquad \qquad    - \Pi^{- -}(x,0,t_0,t_1,t_2) \Big] \ .
\end{align}
In this equation, $\alpha$ denotes the lesser component ($-$) below the Fermi level, $\omega_\text{f} < 0$, and the greater component ($+$) above the Fermi level $\omega_\text{f} > 0$. In addition, we use the abbreviation 
$\Pi^{\beta\gamma}(x,t_0,t_1,t_2,t_3) = G^{\beta}(x,t_0-t_3)G^{\gamma}(x,t_1 - t_2)/G^{\beta}(x,t_1-t_3)G^{\gamma}(x,t_0- t_2)$, 
and the distance $x=x_\text{f}-x_\text{i}$ describes the spatial separation between the emitter and detector along the channel.

\section{Evaluation of low energy excitations for the two-velocity model}

To focus on the energetic sector of Fermi sea excitations centered around the chemical potential $\mu = 0$ for high-energy injected electrons ($\omega_{\text{i}} \gg \bar v / \lambda$), we will neglect the poles of the Green functions in Eq.~(\ref{eq:W}) that contribute to terms decaying exponentially with $\omega_{\text{i}}$. This approach leads to the following expression for both models:
\begin{align}
\label{eq:FSeaExcitations}
	&p_{\text{FS}}(x,\omega_{\text{f}}) =-i\frac{v^2}{2\pi}\int_{-\infty}^\infty\!dt_0\int_{-\infty}^\infty\!dt_1\, e^{i\omega_\mathrm{f}(t_0-t_1)}\nonumber\\
	&\times G^\alpha_\mathrm{c}(0,t_1-t_0)\,\mathrm{exp}[S^+(x,t_0)-S^-(x,t_0)]\nonumber\\
	&\times \mathrm{exp}[S^-(x,t_1)-S^+(x,t_1)] \ .
\end{align}

As a first test for the validity of Eq.~(\ref{eq:FSeaExcitations}),  we apply it to the Green functions of the two-velocity model specified in Eq.~(\ref{eq:Gvbarv}). This method leads to the  following result for the density of  electrons above the Fermi sea ($\omega_{\text{f}}>0$) 
\begin{align}
\label{eq:pTwoVabove}
	& p^{\text{e}}_{\text{FS,2v}} (x_{\text{s}}, \omega_{\text{f}}) = \nonumber \\
	& + \frac{\lambda_{\text{c}}}{\bar{v}} \frac{x_s^2 + \lambda_{\text{c}}^2 \left( 1 - \frac{\bar v}{v} \right)^2}{x_s^2 + \lambda_{\text{c}}^2 \left( 1 + \frac{\bar v}{v} \right)^2}   \left(\frac{4v}{3\bar v} + \frac{2}{3}\right) \exp \left( - 2 \omega_{\text{f}}  \frac{\lambda_{\text{c}}}{\bar v} \right) \ . 
\end{align}
Similarly, we obtain the density of  holes below the Fermi sea ($\omega_{\text{f}}<0$) as
\begin{align}
\label{eq:pTwoVbelow}
	& p^{\text{h}}_{\text{FS,2v}}(x_{\text{s}}, \omega_{\text{f}}) = \nonumber \\
	&- \frac{\lambda_{\text{c}}}{\bar v} \frac{x_s^2 + \lambda_{\text{c}}^2\left(1-\frac{\bar v}{v}\right)^2}{ x_s^2 + \lambda_{\text{c}}^2\left( 1 +\frac{\bar v}{v} \right)^2}\frac{6}{\left( 2  +  \frac{v}{\bar  v}  \right)}  \exp \left( + 2\omega_{\text{f}}\frac{\lambda_{\text{c}}}{v} \right) \ .
\end{align}

\begin{figure}
	\includegraphics[width=0.9\textwidth]{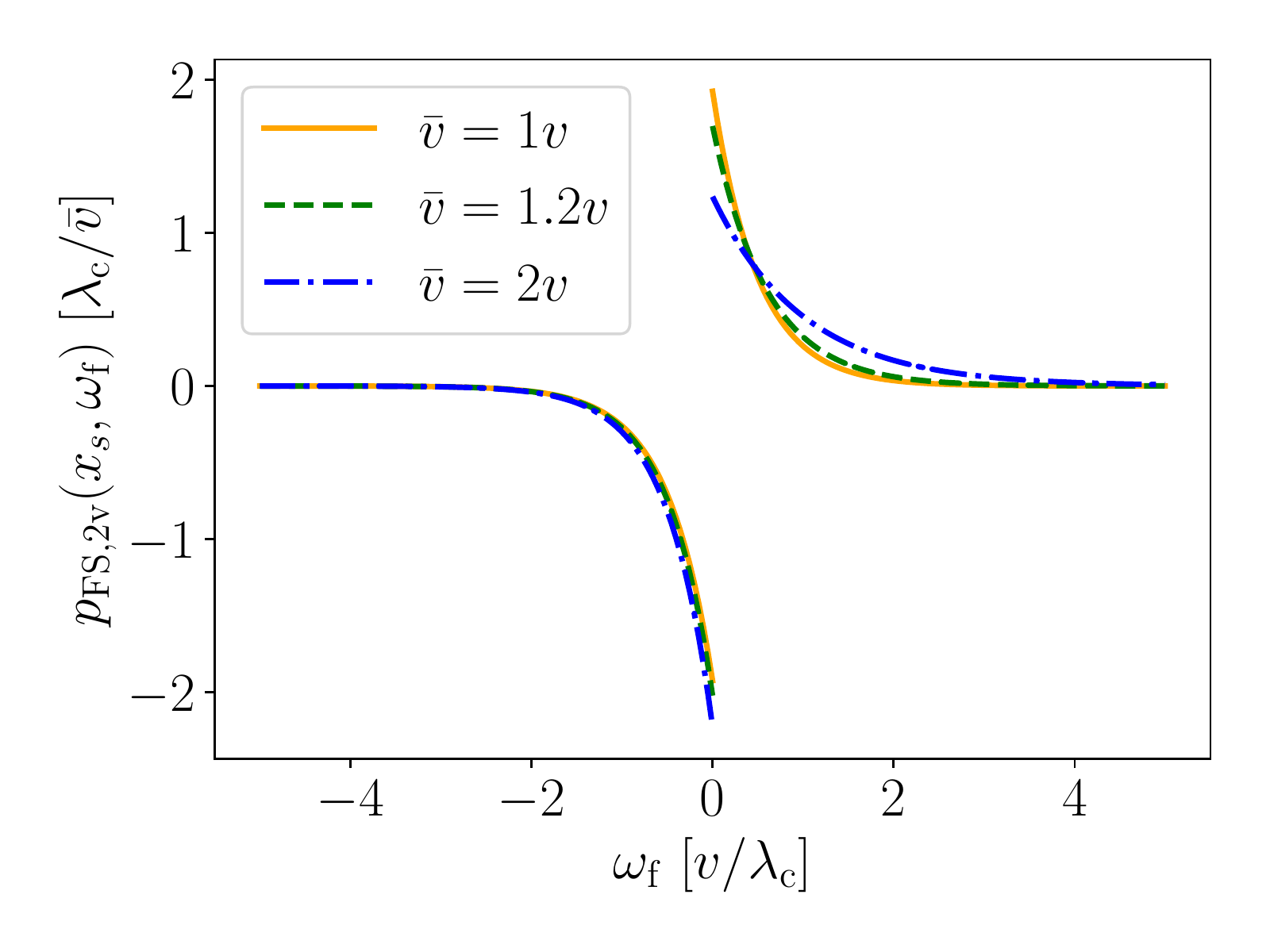}
	\caption{\label{fig:2v}
 Fermi-sea distribution for the two-velocity model, evaluated at $x_s=10\lambda_{\text{c}}$. The resulting distribution, as apparent from  Eqs. (\ref{eq:pTwoVabove}) and (\ref{eq:pTwoVbelow}), decays exponentially as a function of the detection energy. 
 While the distribution is symmetric between positive and negative energies in the scaling limit $\bar{v}=v$ [cf. Eq.~(\ref{eq:p_FS_2v_scaling})], for $\bar{v}>v$ the distribution develops an asymmetry with a slower decay for $\omega_{\text{f}}>0$, but a smaller initial value for $\omega_{\text{f}}=0^+$ as compared to $\omega_{\text{f}}=0^-$. 
}
\end{figure}

Here, $x_{\text{s}} = (\bar v - v)x / v$ corresponds to the spatial dispersion of the wave packet when observed at the detection point~\cite{Fischer.2021}. The expressions in Eqs.~(\ref{eq:pTwoVabove}) and~(\ref{eq:pTwoVbelow}), displayed for $x_{\text{s}} = 10\lambda_{\text{c}}$ in Fig.~\ref{fig:2v}, agree with the results derived for the corresponding Fermi sea excitations when fully evaluating Eq.~(\ref{eq:W}) for the two-velocity model [Eq.~(\ref{eq:Gvbarv})]. This agreement holds true in the limit of high injection energy limit, i.e., when $\omega_{\text{i}} \gg \bar v / \lambda$~\cite{Fischer.2021}.

\section{Scaling limit and evaluation of Fermi sea excitations for exponential interaction}\label{sec:exp_int}

We set out to evaluate the distribution in Eq.~(\ref{eq:FSeaExcitations}) numerically for the full interaction exponent specified in Eq.~(\ref{eq:Sexplicit}). This equation is the outcome of employing the exponential form of interactions found in Eq.~(\ref{eq:nuq}). To achieve this, we consider a scaling limit characterized by large propagation distances ($x \gg \lambda$) and weak interactions ($\nu/2\pi \ll v$). We maintain the product of these quantities, denoted by $x_{\text{s}} = x \nu / 2\pi v$, which corresponds to the spatial dispersion at the detection point, as constant~\cite{Fischer.2021}. The two-velocity solution [Eqs.~(\ref{eq:pTwoVabove}) and~(\ref{eq:pTwoVbelow})] in this limit exhibits symmetry between electrons and holes, expressed as
\begin{equation}\label{eq:p_FS_2v_scaling}
	p_\mathrm{FS,2v}(\omega_\mathrm{f}, x_s)=\alpha\frac{2\lambda_c}{v}\frac{x_s^2}{x_s^2+4\lambda_c^2}e^{-\alpha\omega_\mathrm{f}\frac{2\lambda_c}{v}} . 
\end{equation}
Here, $\alpha=+1$ when above the Fermi sea ($\omega_{\text{f}}>0$) and $\alpha=-1$ when below the Fermi sea ($\omega_{\text{f}}<0$). Notably, the distribution just decays exponentially as a function of the detection energy $\omega_{\text{f}}$ for all $x_s$ values.

For the full model's evaluation, we begin by recognizing that the Green functions independent of $x$ in Eq.~(\ref{eq:FSeaExcitations}) become non-interacting within the scaling limit. By shifting $\tilde{t}_0\to \tilde{t}_0+x/v$ and $\tilde{t}_1\to \tilde{t}_1+x/v$ in Eq.~(\ref{eq:FSeaExcitations}), the scaling limit can be applied to the $x$-dependent exponents, yielding
\begin{equation}\label{eq:S_scaling}
	S^{\text{(exp)}\pm}(x,t+\frac{x}{v})\xrightarrow[\mathrm{limit}]{\mathrm{scaling}}{\sum_{n=1}^\infty \frac{1}{n} \frac{(-x_s)^n}{(vt\mp i\lambda n)^n}} \ .
\end{equation}

By understanding the singularities of the integrand in Eq.~(\ref{eq:FSeaExcitations}), we are able to express the distribution as the limit of a sequence of contour integrals. The integrand of each element of the sequence is proportional to the integrand in Eq.~(\ref{eq:FSeaExcitations}) but with the sums in the exponent restricted to the first $N$ terms.  As a result, for each element of the sequence, the integrand has only a finite number of singularities. The integration contour we consider is given by a semicircle that follows the real line and is closed in the suitable half-plane, encapsulating all singularities within that half-plane. The contribution of the arc segment of the contour decreases in accordance with Jordan's Lemma. For each sequence element, we can then separate the two contour integrals by rewriting the non-interacting Green function as an integral
\begin{equation}\label{eq:integral_for_free_G}
	\frac{1}{v(\tilde{t}_0-\tilde{t}_1)\pm i\epsilon}=\mp iv\int_{0}^{\infty}d\omega\, e^{\pm i\omega((\tilde{t}_0-\tilde{t}_1)\pm i\frac{\epsilon}{v})},
\end{equation}
and then exchanging the contour integrals with the frequency integral. Finally, we can identify the decoupled contour integrals as complex conjugates of each other and thus express the distribution as
\begin{align}\label{eq:Wsl}
p_{\text{FS}}&(x_{\text{s}},\omega_{\text{f}})\xrightarrow[\mathrm{limit}]{\mathrm{scaling}}\lim_{N\to\infty}\frac{\alpha}{4\pi^2}\int_0^\infty\!d\omega\, \left|\int_{C_N}\!dte^{i(|\omega_\mathrm{f}|+\omega)t}\right.\nonumber\\
\times\mathrm{ex}&\mathrm{p}\left.\left(\sum_{n=1}^N \frac{(-x_s)^n}{n}\left[\frac{1}{(vt- i\alpha\lambda n)^n}-\frac{1}{(vt+ i\alpha\lambda n)^n}\right]\right)\right|^2,
\end{align}
where $C_N$ can be any bounded contour enclosing all $N$ singularities in the upper half plane, and $\alpha=+1$ for the electron distribution $(\omega_{\text{f}}>0)$ and $\alpha=-1$ for the hole distribution $(\omega_{\text{f}}<0)$. 
This approach facilitates an efficient numerical computation of excited charge carriers' distributions, extending to values of $x_{\text{s}}$ up to $100\lambda$. In our numerical calculations, we employ dimensionless variables $\xi_\text{s}=x_\text{s}/\lambda$ and $\epsilon_\text{f}=\omega_\text{f}\lambda/v$.

\begin{figure}
	\includegraphics[width=0.95\linewidth]{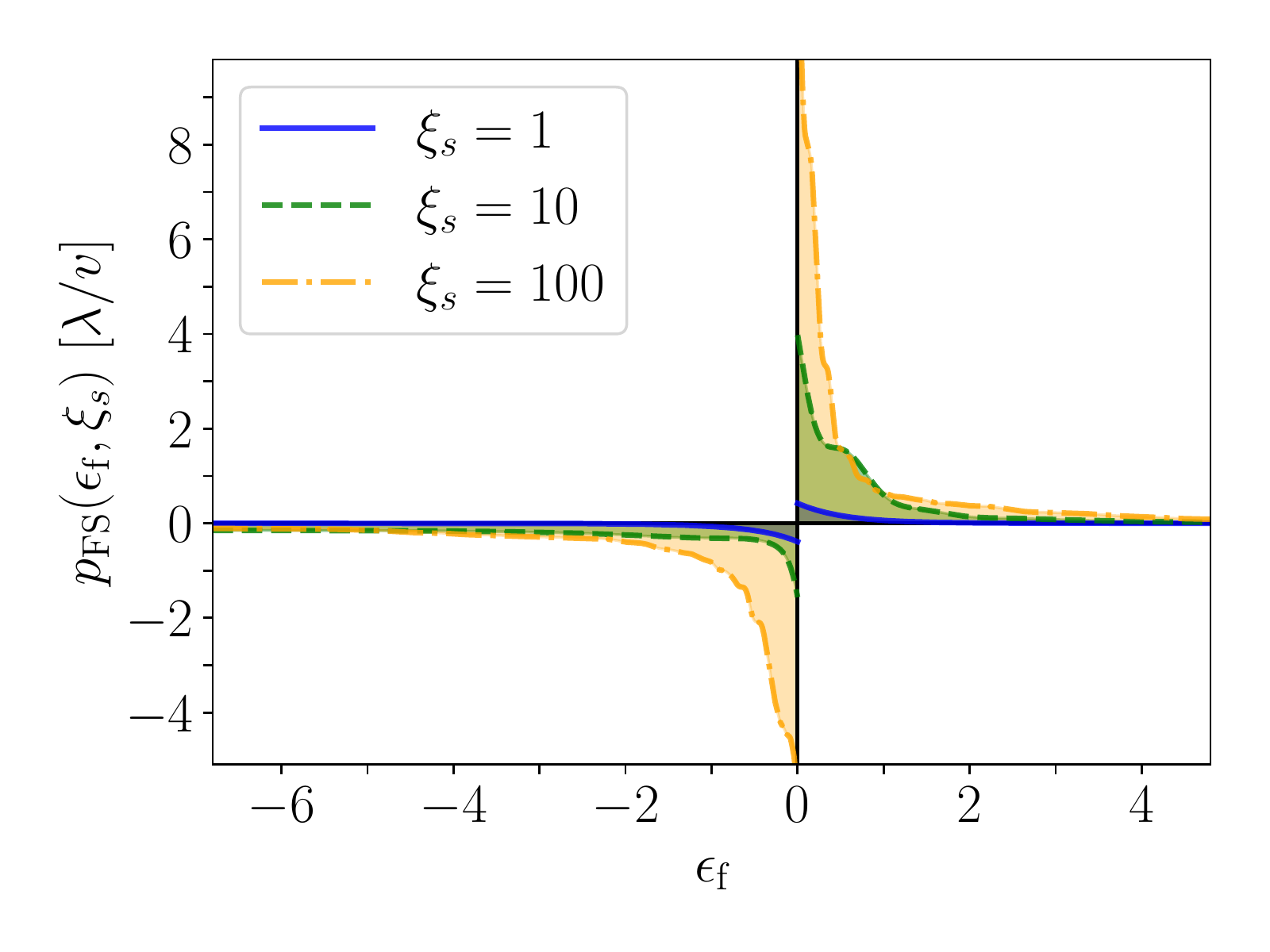}		
	\caption{\label{fig:p_FS} 
	Density $p_\mathrm{FS}$ of Fermi sea excitations for a range of $\xi_s$ values. Interestingly, the distribution of excited electrons ($\epsilon_\mathrm{f}>0$) displays greater concentration near the chemical potential at $\epsilon_\mathrm{f}=0$ compared to the distribution of holes ($\varepsilon_\mathrm{f}<0$).}
\end{figure}

The distributions we have computed are presented in Fig.~\ref{fig:p_FS} for different values of $x_\text{s}$. It is apparent from Eq.~(\ref{eq:Wsl}) that unlike the two-velocity model solution in the scaling limit, given in Eq.~(\ref{eq:p_FS_2v_scaling}), the distributions of electrons and holes are not the same. This is due to the presence of $\alpha$ in the exponent.  As shown in Fig.~\ref{fig:p_FS}, an asymmetry develops between these distributions for larger $x_\text{s}$ values. Contrary to the two-velocity model, the decay is not exponential as a function of detection energy. Instead, the hole distribution starts off with a slow decay, whereas the electron distribution has a higher peak at low energies before rapidly decaying.

\begin{figure}
	\includegraphics[width=.95\linewidth]{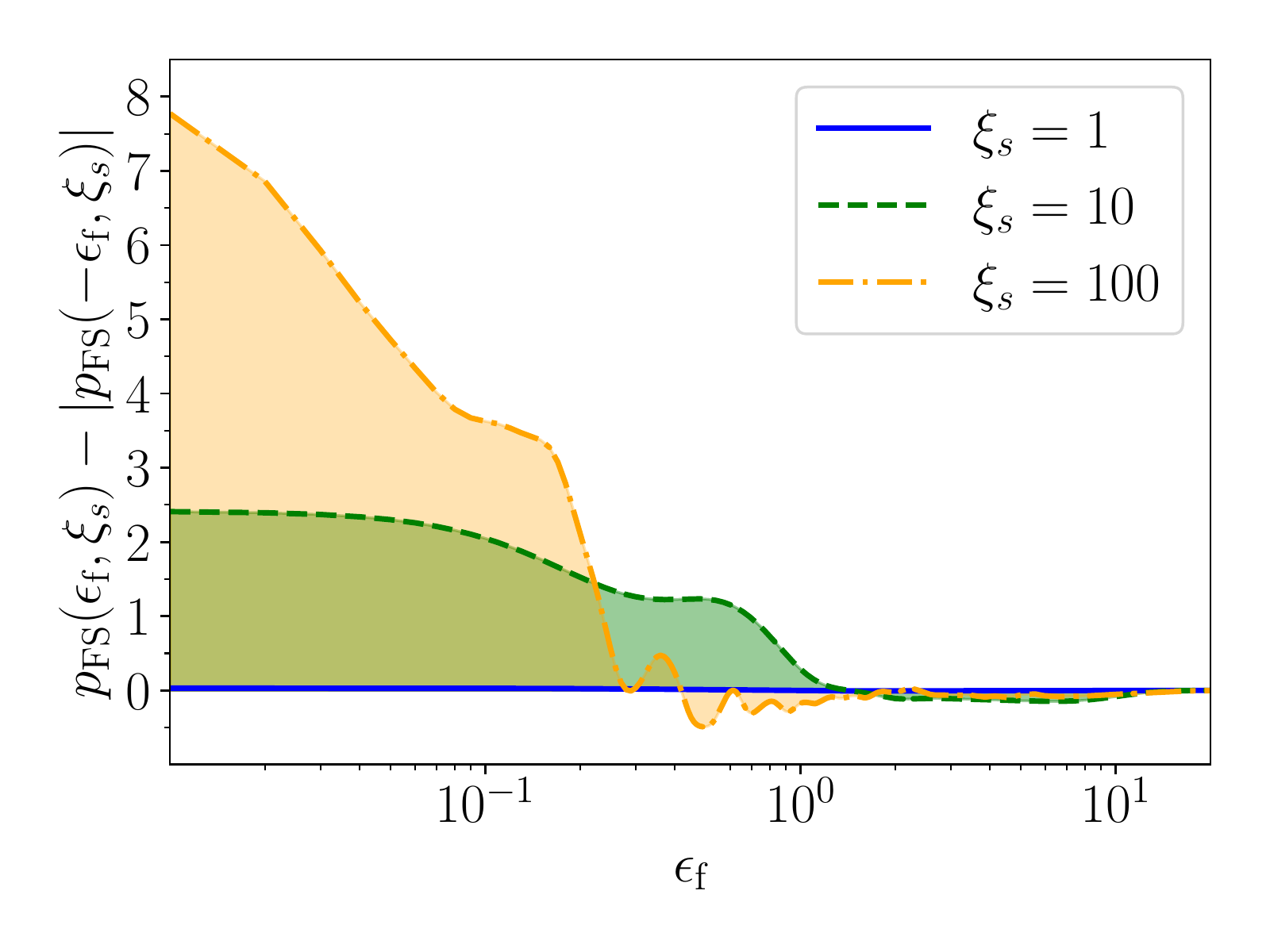}		
	\caption{\label{fig:p_FS_diff} The difference of Fermi sea excitation densities $p_\text{FS}$ for electrons and holes, plotted for the same values of $\xi_s$ in Fig.~\ref{fig:p_FS}. For values $\xi_{\text s}\gg 1$, this difference becomes significant.}
\end{figure}

This asymmetry is visualized in Fig.~\ref{fig:p_FS_diff}, which displays the difference, $|p_\text{FS}(\omega_\text{f})|-|p_\text{FS}(-\omega_\text{f})|$, between the absolute values of the electron and hole distributions at the corresponding $\xi_\text{s}$-values. Notably, the distribution for $\xi_\text{s}=1$ is almost symmetric. Close to the Fermi energy, it only deviates by a few percent from the two-velocity solution given in Eq.~(\ref{eq:p_FS_2v_scaling}). This similarity to the two-velocity solution is even more pronounced for smaller $\xi_s$ values.

By expanding Eq.~(\ref{eq:FSeaExcitations}) for the exponential interaction model in the scaling limit in $x_\text{s}$ and evaluating the integrals up to a specific order in $x_\text{s}$, we gain insight into the similarity between the two solutions for small values of $\xi_\text{s}$. Numerically, we discover that the distribution is perfectly symmetric up to order $(x_\text{s}/\lambda)^2$. Any asymmetric contributions are at least of order $(x_\text{s}/\lambda)^4$. Consequently, the total distribution is nearly symmetric in the regime of weak interaction strength and short propagation length ($x_\text{s}/\lambda\leq 1$), which second-order perturbation theory adequately captures.

Additionally, we analytically derived the lowest non-vanishing order term of $(x_\text{s}/\lambda)^2$ and found that it matches the $(x_\text{s}/\lambda)^2$ term obtained from expanding Eq.~(\ref{eq:p_FS_2v_scaling}) for small $x_{\text s}$ when $\lambda_\text{c}=\lambda$. This clarifies why the two-velocity solution agrees well with the solution from the exponential interaction model for small $x_\text{s}/\lambda$ values.

\section{Conclusions}

In this study, we have explored the excitation dynamics of charge carriers originating from the Fermi sea within a one-dimensional chiral channel,  driven out of equilibrium by the dissipated energy of high-energy injected electrons. We focused on  Fermi sea excitations by discarding terms that decay exponentially with the injection energy. We evaluated the robustness of this approach by juxtaposing the expression obtained by neglecting the exponentially decaying terms with the result of a full calculation of the channel's distribution and then taking the limit of high injection energy. This comparison was conducted within an analytical model that contains only two different excitations velocities.  

Upon performing a numerical analysis of the densities of charge carriers excited from the Fermi sea for the full model 
within a scaling limit, we discovered an asymmetry between the excitations of electrons and holes. This asymmetry becomes apparent when the propagation distances and interaction strengths surpass the applicability of second-order perturbation theory.

\appendix
\section{Particle number and energy conservation}

\begin{figure}
	\centering
	\includegraphics[width=0.92\linewidth]{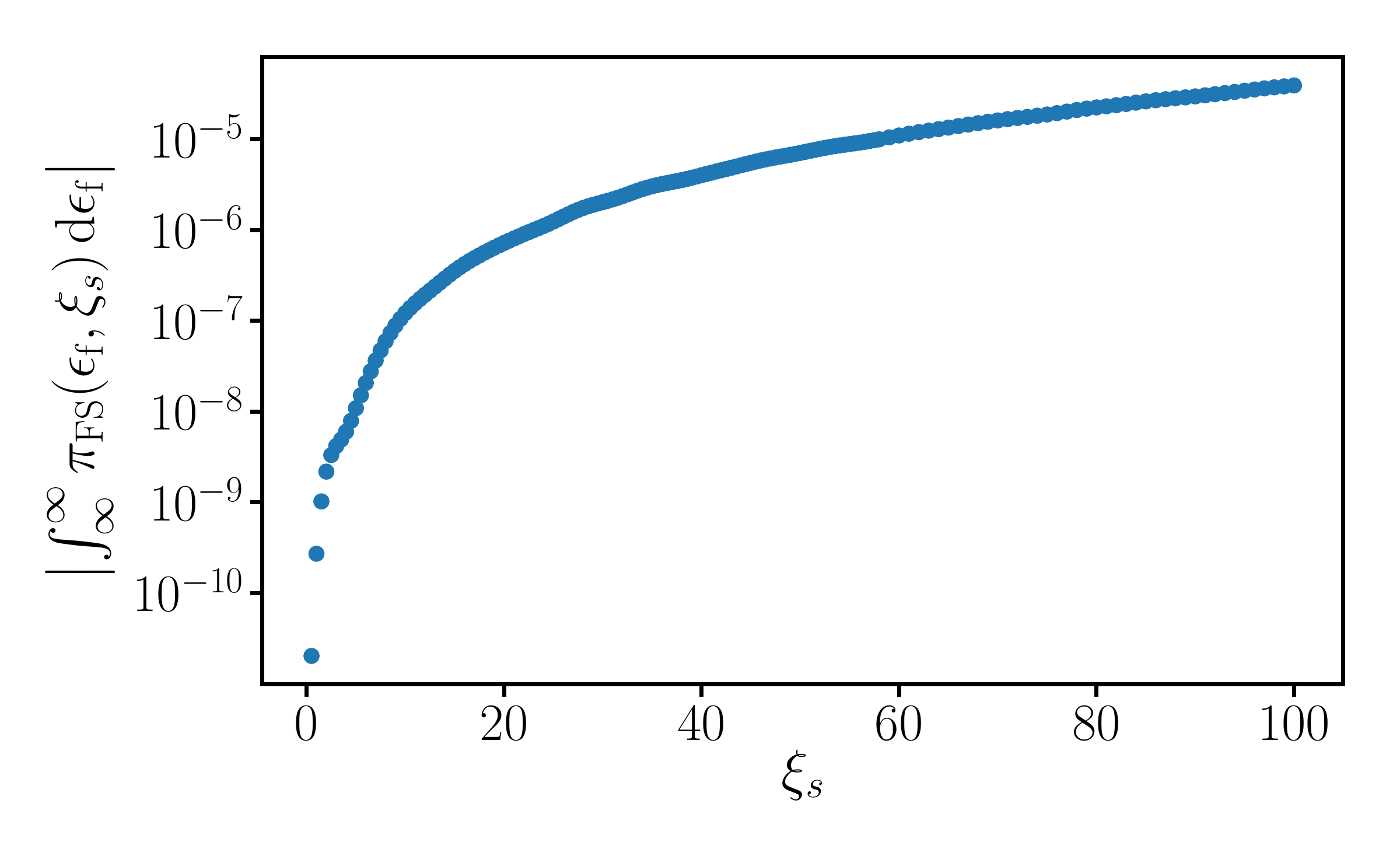}		
	\caption{\label{fig:Gewichte}   Numerical verification of particle number conservation within the Fermi-sea excitation distribution. We find that particle number conservation is satisfied for all evaluated values of $\xi_s$.}
\end{figure}

To validate the numerically obtained densities of excited charge carriers, we will look at the conservation of both particle number and energy of the relevant components. In this context, particle number is conserved separately for the electron density $p_{\text{HE}}$ at high injection energy, as described in Ref.~\cite{Fischer.2021}, and for Fermi sea excitations $p_{\text{FS}}$. The former quantity is represented in a dimensionless form as $\pi_{\text{HE}} = p_{\text{HE}} v / \lambda$, fulfilling $\int_0^\infty d\epsilon_{\text{if}}\,  \pi_{\text{HE}} (\epsilon_{\text{if}}) = 1$, where $\epsilon_{\text{if}}$ is the dimensionless energy loss $\epsilon_{\text{i}}-\epsilon_{\text{f}}$. Consequently, we anticipate $\int_{-\infty}^{\infty} d\epsilon_\text{f} \, \pi_{\text{FS}} (\epsilon_{\text{f}}) = 0$, a condition that holds true for all $\xi_{\text{s}} = x_{\text{s}} / \lambda$ within the bounds of numerical accuracy, as shown in Fig.~\ref{fig:Gewichte}.

\begin{figure}
	\centering
	\includegraphics[width=\linewidth]{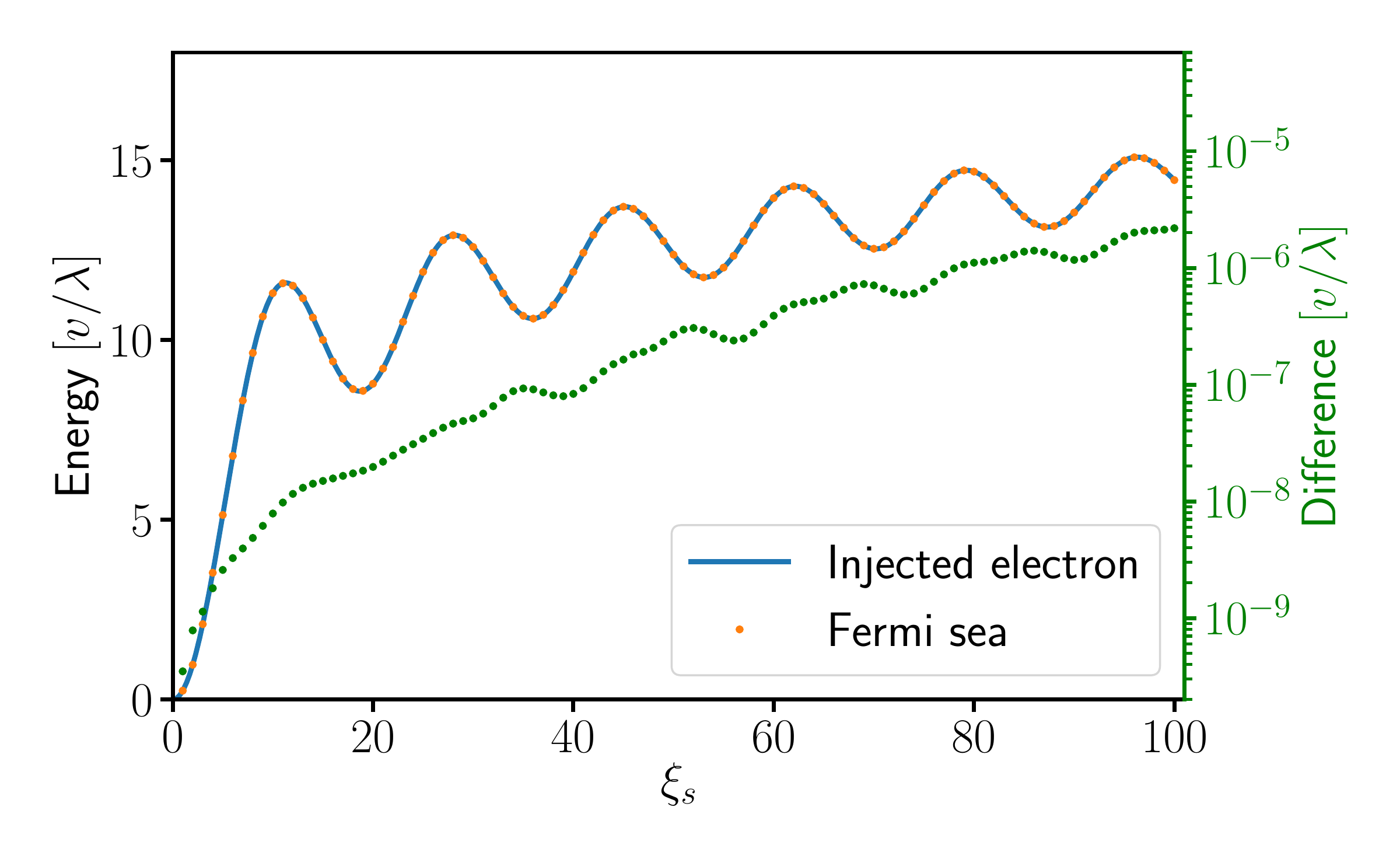}	
	\caption{\label{fig:energy_cont} Numerically computed energy gain in the Fermi sea $\epsilon_{\text{G}}$ and the energy dissipated by the injected electron $\epsilon_{\text{D}}$. The green dots shows the difference between $\epsilon_{\text{G}}$ and $\epsilon_{\text{D}}$. The difference is due to numerical errors.}
\end{figure}

In addition, we expect that the energy dissipated by the injected electron, defined as $\epsilon_{\text{D}}=\int_0^\infty d\epsilon_{\text{if}} \, \epsilon_{\text{if}} \, \pi_{\text{HE}} (\epsilon_{\text{if}})$, is equal to the energy gain of the Fermi sea, represented as $\epsilon_{\text{G}} = \int_{-\infty}^{\infty} d\epsilon_f \, \epsilon_{\text{f}} \, \pi_{\text{FS}} (\epsilon_{\text{f}})$. The energy dissipated by the injected electron can be represented as a series,
\begin{equation}\label{eq:E_HE}
\epsilon_\text{D}=\sum_{n=1}^\infty \frac{(-1)^{n-1}}{n}\left(\frac{\xi_s}{2n}\right)^{2n} \ .
\end{equation}
By comparing the numerically obtained results for $\epsilon_{\text{G}}$ and $\epsilon_{\text{D}}$ using Eq.~(\ref{eq:E_HE}), we confirm that these quantities coincide within the bounds of numerical accuracy, as depicted in Fig.~\ref{fig:energy_cont}.

\section{On the asymmetry of the distribution}
\begin{figure}[h]
	\centering
	\includegraphics[width=0.99\linewidth]{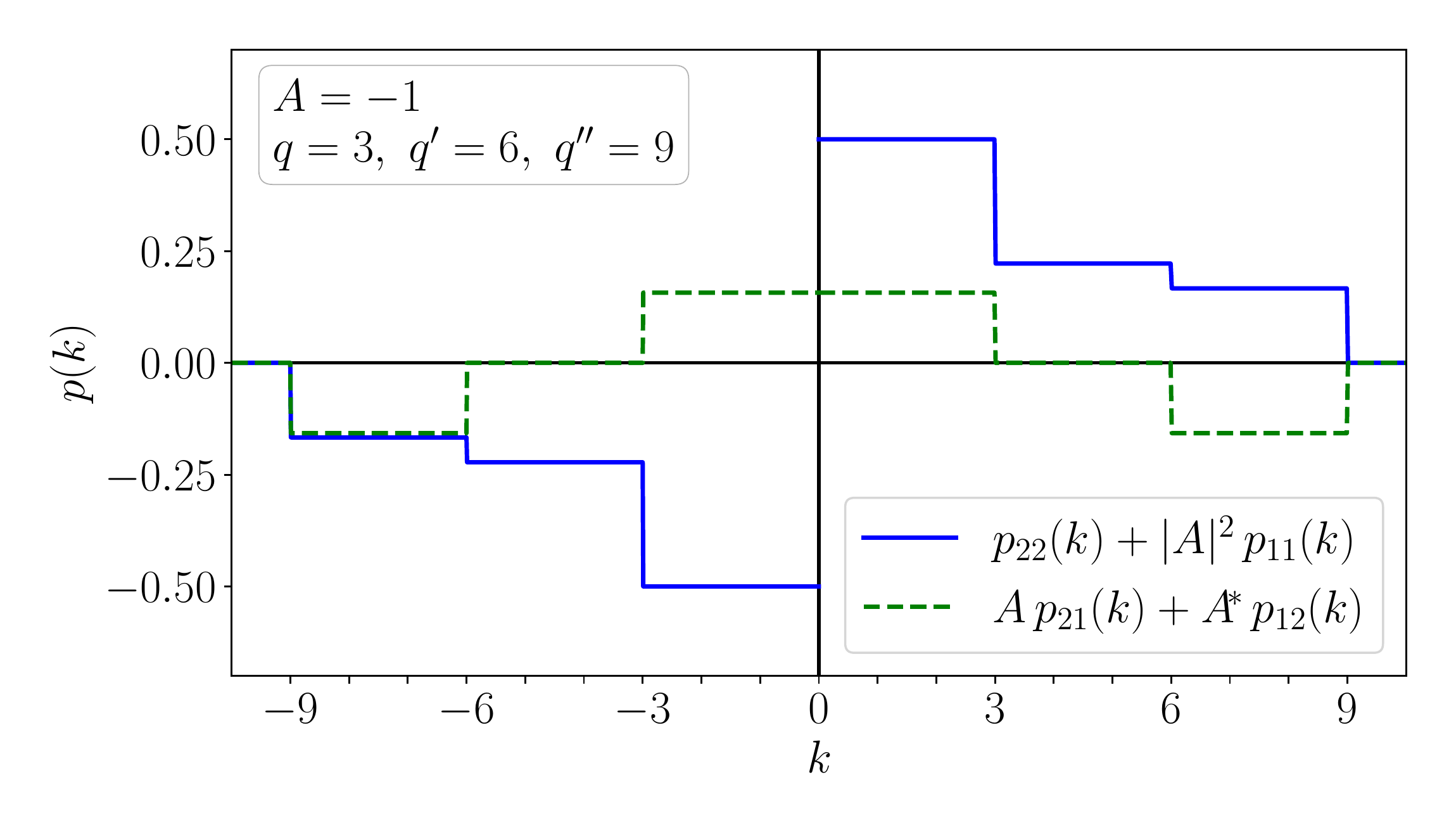}
	\caption{The even and the odd contribution to the distribution of excited electrons and holes in the state $|q,q^\prime,q^{\prime\prime}\rangle$ in Eq.~(\ref{eq:mix_state}) plotted for $A=-1,q=3,q^\prime=6,q^{\prime\prime}=9$. Here, we neglect the dimension of $q$, since it cancels with the momentum integrals in the continuum limit and the dimension of $k$ since it enters only as an index of the fermion operators.}\label{fig:2ps}
\end{figure}

To investigate the origin of the asymmetry of electron and hole excitations, we note that the interaction Hamiltonian  [cf.~second term of Eq.~(\ref{eq:H})] in the limit of high injection energy can be bosonized in the form~\cite{Fischer.2021,Heyl.2010},
\begin{align}
	\label{eq:HhighE}
	H_V = -i \int_{-\infty}^\infty dk \int_0^{\infty}dq \sqrt{q} \frac{\nu_q}{2\pi} \left( c_{k-q}^{\dagger}c_k^{\vphantom{\dagger}} b_{q}^{\dagger} - c_{k+q}^{\dagger}c_k^{\vphantom{\dagger}} b_{q}^{\vphantom{\dagger}} \right),
\end{align}
where the injected electron (described by the fermion operators $c_k^{\vphantom{\dagger}},c_k^\dagger$) can be treated separately from plasmons in the Fermi sea (described by boson operators $b_q^{\vphantom{\dagger}},b_q^\dagger$). This form of the Hamiltonian is not invariant under the particle-hole conjugation operation for the fermion operators. 
As a consequence of energy relaxation, in the Fermi sea, we expect a mixture of states with different numbers of plasmons. As an example, we consider the simplest mixture of a single plasmon state and a state with two plasmons 
\begin{equation}\label{eq:mix_state}
	|q,q^\prime,q^{\prime\prime}\rangle=[b_{q}^\dag b_{q^\prime}^\dag + A\cdot b_{q^{\prime\prime}}^\dag]|0\rangle \, .
\end{equation}
For this state, we  compute the expectation value of the electron occupation relative to the equilibrium ground state 
\begin{equation}\label{eq:n_k}
	\langle n(k)\rangle=\langle q,q^\prime,q^{\prime\prime}|\phantom{.}^*_*c^\dagger_k{c_k}^*_*|q,q^\prime,q^{\prime\prime}\rangle \, ,
\end{equation}
where $\phantom{.}^*_*c^\dagger_k{c_k}^*_*=c^\dagger_k{c_k}-\langle 0|c^\dagger_k{c_k}|0\rangle$ is the occupation relative to the equilibrium ground state $|0\rangle$. The expectation value yields the density of excited electrons. For momenta smaller than the Fermi momentum the density is negative, which corresponds to a positive density of holes. 
 The expectation value in Eq. (\ref{eq:n_k})
splits up into four parts, namely, the two expectation values in the one- or two-plasmon state, respectively,
\begin{align}\label{eq:p_xx}
	p&_{22}(k;q,q^\prime)+|A|^2\cdot p_{11}(k;q^{\prime\prime})=\nonumber\\
	&\langle 0| b_{q^\prime}b_q \vphantom{a}_{*}^{*}c_k^\dag c_k\vphantom{a}_{*}^{*} b^\dag_q b^\dag_{q^\prime}|0\rangle+|A|^2\cdot\langle 0| b_{q^{\prime\prime}} \vphantom{a}_{*}^{*}c_k^\dag c_k\vphantom{a}_{*}^{*}  b^\dag_{q^{\prime\prime}}|0\rangle,
\end{align}
and the two mixing terms
\begin{align}\label{eq:p_xy}
	A\cdot p_{21}(k&;q,q^\prime,q^{\prime\prime})+A^*\cdot p_{12}(k;q,q^\prime,q^{\prime\prime})=\nonumber\\
	A\cdot\langle& 0| b_{q^\prime}b_q \vphantom{a}_{*}^{*}c_k^\dag c_k\vphantom{a}_{*}^{*} b^\dag_{q^{\prime\prime}}|0\rangle
	+A^*\cdot\langle 0| b_{q^{\prime\prime}} \vphantom{a}_{*}^{*}c_k^\dag c_k\vphantom{a}_{*}^{*} b^\dag_q b^\dag_{q^\prime}|0\rangle.
\end{align}
It turns out that the two mixing terms $p_{12}$ and $p_{21}$ are identical, such that $A\cdot p_{21}+A^*\!\cdot p_{12}=2\text{Re}(A)\cdot p_{12}$. When expressing the boson operators in terms of fermion operators,
\begin{equation}\label{eq:boson_fermion_representation}
	b^\dag_q=\sqrt{\frac{2\pi}{Lq}}\sum_{k'}c^\dag_{k'+q}c_{k'},
\end{equation}
we can calculate the expectation value in Eq. (\ref{eq:n_k}) using Wick's theorem and taking the continuum limit $\frac{1}{L}\sum_k\to\int \frac{dk}{2\pi}$. We find that $p_{11}$ and $p_{22}$ are odd functions of $k$ as one would expect. The mixing terms, however, are even functions of $k$, cf. Fig.~\ref{fig:2ps}.

\begin{figure}
	\centering
	\includegraphics[width=.95\linewidth]{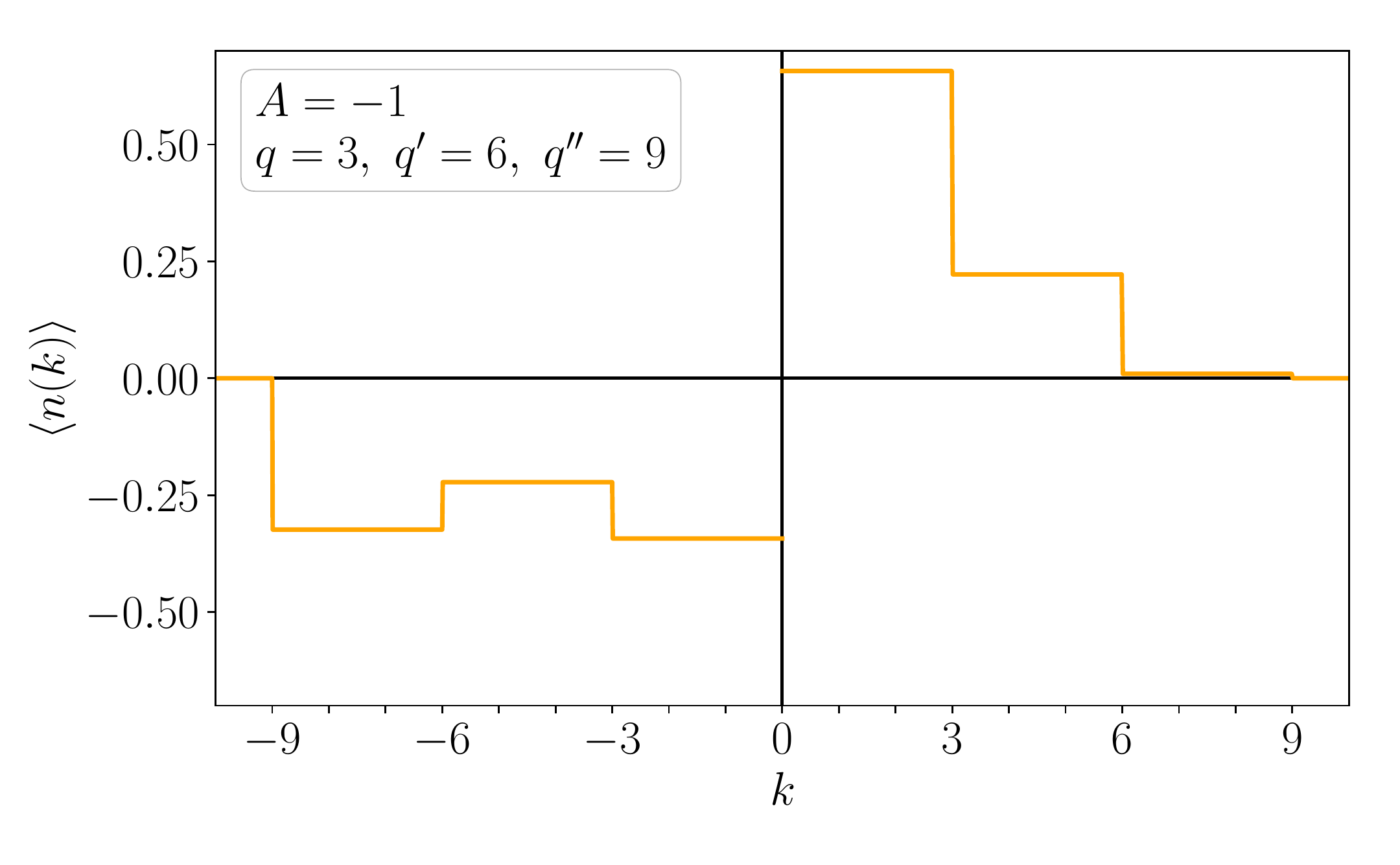}
	\caption{The distribution of excited electrons and holes in the state $|q,q^\prime,q^{\prime\prime}\rangle$ in Eq.~(\ref{eq:mix_state}) plotted for $A=-1,q=3,q^\prime=6,q^{\prime\prime}=9$. The asymmetry for these values is similar to the one that was found in Sec.~\ref{sec:exp_int} for the Fermi-sea distribution $p_\text{FS}$.}\label{fig:p_tot}
\end{figure}

For a negative real part of $A$ we then obtain an asymmetry in the expectation value of the occupation number,  similar to the one we found for the Fermi sea distribution $p_\text{FS}$ in Sec. \ref{sec:exp_int}, cf. Fig.~\ref{fig:p_tot}. For a positive real part of $A$, however, we would get an asymmetry opposite to the one we found in Sec.~\ref{sec:exp_int}. 
This shows that an asymmetry in the distribution of excited electrons and holes can arise when mixing states with a different number of plasmons. An intuitive interpretation of why the particular asymmetry that we found in Sec.~\ref{sec:exp_int}, arises, would be desirable but cannot be obtained from the simple model calculation presented here alone.  

\vspace*{11cm}

\begin{thebibliography}{42}
	\expandafter\ifx\csname natexlab\endcsname\relax\def\natexlab#1{#1}\fi
	\expandafter\ifx\csname bibnamefont\endcsname\relax
	  \def\bibnamefont#1{#1}\fi
	\expandafter\ifx\csname bibfnamefont\endcsname\relax
	  \def\bibfnamefont#1{#1}\fi
	\expandafter\ifx\csname citenamefont\endcsname\relax
	  \def\citenamefont#1{#1}\fi
	\expandafter\ifx\csname url\endcsname\relax
	  \def\url#1{\texttt{#1}}\fi
	\expandafter\ifx\csname urlprefix\endcsname\relax\def\urlprefix{URL }\fi
	\providecommand{\bibinfo}[2]{#2}
	\providecommand{\eprint}[2][]{\url{#2}}
	
	\bibitem[{\citenamefont{Fischer et~al.}(2021)\citenamefont{Fischer, Meir,
	  Gefen, and Rosenow}}]{Fischer.2021}
	\bibinfo{author}{\bibfnamefont{S.~G.} \bibnamefont{Fischer}},
	  \bibinfo{author}{\bibfnamefont{Y.}~\bibnamefont{Meir}},
	  \bibinfo{author}{\bibfnamefont{Y.}~\bibnamefont{Gefen}}, \bibnamefont{and}
	  \bibinfo{author}{\bibfnamefont{B.}~\bibnamefont{Rosenow}}
	  (\bibinfo{year}{2021}), \bibinfo{note}{preprint at:
	  {\textless}https://arxiv.org/abs/2108.00685{\textgreater}}.
	
	\bibitem[{\citenamefont{Polkovnikov et~al.}(2011)\citenamefont{Polkovnikov,
	  Sengupta, Silva, and Vengalattore}}]{Polkovnikov.2011}
	\bibinfo{author}{\bibfnamefont{A.}~\bibnamefont{Polkovnikov}},
	  \bibinfo{author}{\bibfnamefont{K.}~\bibnamefont{Sengupta}},
	  \bibinfo{author}{\bibfnamefont{A.}~\bibnamefont{Silva}}, \bibnamefont{and}
	  \bibinfo{author}{\bibfnamefont{M.}~\bibnamefont{Vengalattore}},
	  \bibinfo{journal}{{Reviews of Modern Physics}} \textbf{\bibinfo{volume}{83}},
	  \bibinfo{pages}{863} (\bibinfo{year}{2011}).
	
	\bibitem[{\citenamefont{Gogolin and Eisert}(2016)}]{Gogolin.2016}
	\bibinfo{author}{\bibfnamefont{C.}~\bibnamefont{Gogolin}} \bibnamefont{and}
	  \bibinfo{author}{\bibfnamefont{J.}~\bibnamefont{Eisert}},
	  \bibinfo{journal}{{Reports on progress in physics. Physical Society (Great
	  Britain)}} \textbf{\bibinfo{volume}{79}}, \bibinfo{pages}{056001}
	  (\bibinfo{year}{2016}).
	
	\bibitem[{\citenamefont{Milletar{\`i} and Rosenow}(2013)}]{Milletari.2013}
	\bibinfo{author}{\bibfnamefont{M.}~\bibnamefont{Milletar{\`i}}}
	  \bibnamefont{and} \bibinfo{author}{\bibfnamefont{B.}~\bibnamefont{Rosenow}},
	  \bibinfo{journal}{{Physical Review Letters}} \textbf{\bibinfo{volume}{111}},
	  \bibinfo{pages}{136807} (\bibinfo{year}{2013}).
	
	\bibitem[{\citenamefont{Schneider et~al.}(2017)\citenamefont{Schneider,
	  Milletari, and Rosenow}}]{Schneider.2017}
	\bibinfo{author}{\bibfnamefont{A.}~\bibnamefont{Schneider}},
	  \bibinfo{author}{\bibfnamefont{M.}~\bibnamefont{Milletari}},
	  \bibnamefont{and} \bibinfo{author}{\bibfnamefont{B.}~\bibnamefont{Rosenow}},
	  \bibinfo{journal}{{SciPost Physics}} \textbf{\bibinfo{volume}{2}},
	  \bibinfo{pages}{007} (\bibinfo{year}{2017}).
	
	\bibitem[{\citenamefont{Kinoshita et~al.}(2006)\citenamefont{Kinoshita, Wenger,
	  and Weiss}}]{Kinoshita.2006}
	\bibinfo{author}{\bibfnamefont{T.}~\bibnamefont{Kinoshita}},
	  \bibinfo{author}{\bibfnamefont{T.}~\bibnamefont{Wenger}}, \bibnamefont{and}
	  \bibinfo{author}{\bibfnamefont{D.~S.} \bibnamefont{Weiss}},
	  \bibinfo{journal}{{Nature}} \textbf{\bibinfo{volume}{440}},
	  \bibinfo{pages}{900} (\bibinfo{year}{2006}).
	
	\bibitem[{\citenamefont{Hofferberth et~al.}(2007)\citenamefont{Hofferberth,
	  Lesanovsky, Fischer, Schumm, and Schmiedmayer}}]{Hofferberth.2007}
	\bibinfo{author}{\bibfnamefont{S.}~\bibnamefont{Hofferberth}},
	  \bibinfo{author}{\bibfnamefont{I.}~\bibnamefont{Lesanovsky}},
	  \bibinfo{author}{\bibfnamefont{B.}~\bibnamefont{Fischer}},
	  \bibinfo{author}{\bibfnamefont{T.}~\bibnamefont{Schumm}}, \bibnamefont{and}
	  \bibinfo{author}{\bibfnamefont{J.}~\bibnamefont{Schmiedmayer}},
	  \bibinfo{journal}{{Nature}} \textbf{\bibinfo{volume}{449}},
	  \bibinfo{pages}{324} (\bibinfo{year}{2007}).
	
	\bibitem[{\citenamefont{Lerner et~al.}(2008)\citenamefont{Lerner, Yudson, and
	  Yurkevich}}]{Lerner.2008}
	\bibinfo{author}{\bibfnamefont{I.~V.} \bibnamefont{Lerner}},
	  \bibinfo{author}{\bibfnamefont{V.~I.} \bibnamefont{Yudson}},
	  \bibnamefont{and} \bibinfo{author}{\bibfnamefont{I.~V.}
	  \bibnamefont{Yurkevich}}, \bibinfo{journal}{{Physical Review Letters}}
	  \textbf{\bibinfo{volume}{100}}, \bibinfo{pages}{256805}
	  (\bibinfo{year}{2008}).
	
	\bibitem[{\citenamefont{Calzona et~al.}(2017)\citenamefont{Calzona, Gambetta,
	  Carrega, Cavaliere, and Sassetti}}]{Calzona.2017}
	\bibinfo{author}{\bibfnamefont{A.}~\bibnamefont{Calzona}},
	  \bibinfo{author}{\bibfnamefont{F.~M.} \bibnamefont{Gambetta}},
	  \bibinfo{author}{\bibfnamefont{M.}~\bibnamefont{Carrega}},
	  \bibinfo{author}{\bibfnamefont{F.}~\bibnamefont{Cavaliere}},
	  \bibnamefont{and} \bibinfo{author}{\bibfnamefont{M.}~\bibnamefont{Sassetti}},
	  \bibinfo{journal}{{Physical Review B}} \textbf{\bibinfo{volume}{95}},
	  \bibinfo{pages}{085101} (\bibinfo{year}{2017}).
	
	\bibitem[{\citenamefont{Calzona et~al.}(2018)\citenamefont{Calzona, Gambetta,
	  Carrega, Cavaliere, Schmidt, and Sassetti}}]{Calzona.2018b}
	\bibinfo{author}{\bibfnamefont{A.}~\bibnamefont{Calzona}},
	  \bibinfo{author}{\bibfnamefont{F.~M.} \bibnamefont{Gambetta}},
	  \bibinfo{author}{\bibfnamefont{M.}~\bibnamefont{Carrega}},
	  \bibinfo{author}{\bibfnamefont{F.}~\bibnamefont{Cavaliere}},
	  \bibinfo{author}{\bibfnamefont{T.}~\bibnamefont{Schmidt}}, \bibnamefont{and}
	  \bibinfo{author}{\bibfnamefont{M.}~\bibnamefont{Sassetti}},
	  \bibinfo{journal}{{SciPost Physics}} \textbf{\bibinfo{volume}{4}},
	  \bibinfo{pages}{023} (\bibinfo{year}{2018}).
	
	\bibitem[{\citenamefont{{\v{S}}trkalj et~al.}(2019)\citenamefont{{\v{S}}trkalj,
	  Ferguson, Wolf, Levkivskyi, and Zilberberg}}]{Strkalj.2019}
	\bibinfo{author}{\bibfnamefont{A.}~\bibnamefont{{\v{S}}trkalj}},
	  \bibinfo{author}{\bibfnamefont{M.~S.} \bibnamefont{Ferguson}},
	  \bibinfo{author}{\bibfnamefont{T.~M.~R.} \bibnamefont{Wolf}},
	  \bibinfo{author}{\bibfnamefont{I.}~\bibnamefont{Levkivskyi}},
	  \bibnamefont{and}
	  \bibinfo{author}{\bibfnamefont{O.}~\bibnamefont{Zilberberg}},
	  \bibinfo{journal}{{Physical Review Letters}} \textbf{\bibinfo{volume}{122}},
	  \bibinfo{pages}{126802} (\bibinfo{year}{2019}).
	
	\bibitem[{\citenamefont{Deshpande et~al.}(2010)\citenamefont{Deshpande,
	  Bockrath, Glazman, and Yacoby}}]{Deshpande.2010}
	\bibinfo{author}{\bibfnamefont{V.~V.} \bibnamefont{Deshpande}},
	  \bibinfo{author}{\bibfnamefont{M.}~\bibnamefont{Bockrath}},
	  \bibinfo{author}{\bibfnamefont{L.~I.} \bibnamefont{Glazman}},
	  \bibnamefont{and} \bibinfo{author}{\bibfnamefont{A.}~\bibnamefont{Yacoby}},
	  \bibinfo{journal}{{Nature}} \textbf{\bibinfo{volume}{464}},
	  \bibinfo{pages}{209} (\bibinfo{year}{2010}).
	
	\bibitem[{\citenamefont{Barak et~al.}(2010)\citenamefont{Barak, Steinberg,
	  Pfeiffer, West, Glazman, von Oppen, and Yacoby}}]{Barak.2010}
	\bibinfo{author}{\bibfnamefont{G.}~\bibnamefont{Barak}},
	  \bibinfo{author}{\bibfnamefont{H.}~\bibnamefont{Steinberg}},
	  \bibinfo{author}{\bibfnamefont{L.~N.} \bibnamefont{Pfeiffer}},
	  \bibinfo{author}{\bibfnamefont{K.~W.} \bibnamefont{West}},
	  \bibinfo{author}{\bibfnamefont{L.}~\bibnamefont{Glazman}},
	  \bibinfo{author}{\bibfnamefont{F.}~\bibnamefont{von Oppen}},
	  \bibnamefont{and} \bibinfo{author}{\bibfnamefont{A.}~\bibnamefont{Yacoby}},
	  \bibinfo{journal}{{Nature Physics}} \textbf{\bibinfo{volume}{6}},
	  \bibinfo{pages}{489} (\bibinfo{year}{2010}).
	
	\bibitem[{\citenamefont{Idrisov and Schmidt}(2019)}]{Idrisov.2019}
	\bibinfo{author}{\bibfnamefont{E.~G.} \bibnamefont{Idrisov}} \bibnamefont{and}
	  \bibinfo{author}{\bibfnamefont{T.~L.} \bibnamefont{Schmidt}},
	  \bibinfo{journal}{{Physical Review B}} \textbf{\bibinfo{volume}{100}},
	  \bibinfo{pages}{165404} (\bibinfo{year}{2019}).
	
	\bibitem[{\citenamefont{Kr{\"a}henmann
	  et~al.}(2019)\citenamefont{Kr{\"a}henmann, Fischer, R{\"o}{\"o}sli, Ihn,
	  Reichl, Wegscheider, Ensslin, Gefen, and Meir}}]{Krahenmann.2019}
	\bibinfo{author}{\bibfnamefont{T.}~\bibnamefont{Kr{\"a}henmann}},
	  \bibinfo{author}{\bibfnamefont{S.~G.} \bibnamefont{Fischer}},
	  \bibinfo{author}{\bibfnamefont{M.}~\bibnamefont{R{\"o}{\"o}sli}},
	  \bibinfo{author}{\bibfnamefont{T.}~\bibnamefont{Ihn}},
	  \bibinfo{author}{\bibfnamefont{C.}~\bibnamefont{Reichl}},
	  \bibinfo{author}{\bibfnamefont{W.}~\bibnamefont{Wegscheider}},
	  \bibinfo{author}{\bibfnamefont{K.}~\bibnamefont{Ensslin}},
	  \bibinfo{author}{\bibfnamefont{Y.}~\bibnamefont{Gefen}}, \bibnamefont{and}
	  \bibinfo{author}{\bibfnamefont{Y.}~\bibnamefont{Meir}},
	  \bibinfo{journal}{{Nature communications}} \textbf{\bibinfo{volume}{10}},
	  \bibinfo{pages}{3915} (\bibinfo{year}{2019}).
	
	\bibitem[{\citenamefont{Rodriguez et~al.}(2020)\citenamefont{Rodriguez,
	  Parmentier, Ferraro, Roulleau, Gennser, Cavanna, Sassetti, Portier, Mailly,
	  and Roche}}]{Rodriguez.2020}
	\bibinfo{author}{\bibfnamefont{R.~H.} \bibnamefont{Rodriguez}},
	  \bibinfo{author}{\bibfnamefont{F.~D.} \bibnamefont{Parmentier}},
	  \bibinfo{author}{\bibfnamefont{D.}~\bibnamefont{Ferraro}},
	  \bibinfo{author}{\bibfnamefont{P.}~\bibnamefont{Roulleau}},
	  \bibinfo{author}{\bibfnamefont{U.}~\bibnamefont{Gennser}},
	  \bibinfo{author}{\bibfnamefont{A.}~\bibnamefont{Cavanna}},
	  \bibinfo{author}{\bibfnamefont{M.}~\bibnamefont{Sassetti}},
	  \bibinfo{author}{\bibfnamefont{F.}~\bibnamefont{Portier}},
	  \bibinfo{author}{\bibfnamefont{D.}~\bibnamefont{Mailly}}, \bibnamefont{and}
	  \bibinfo{author}{\bibfnamefont{P.}~\bibnamefont{Roche}},
	  \bibinfo{journal}{{Nature communications}} \textbf{\bibinfo{volume}{11}},
	  \bibinfo{pages}{2426} (\bibinfo{year}{2020}).
	
	\bibitem[{\citenamefont{Kamata et~al.}(2010)\citenamefont{Kamata, Ota, Muraki,
	  and Fujisawa}}]{Kamata.2010}
	\bibinfo{author}{\bibfnamefont{H.}~\bibnamefont{Kamata}},
	  \bibinfo{author}{\bibfnamefont{T.}~\bibnamefont{Ota}},
	  \bibinfo{author}{\bibfnamefont{K.}~\bibnamefont{Muraki}}, \bibnamefont{and}
	  \bibinfo{author}{\bibfnamefont{T.}~\bibnamefont{Fujisawa}},
	  \bibinfo{journal}{{Physical Review B}} \textbf{\bibinfo{volume}{81}},
	  \bibinfo{pages}{085329} (\bibinfo{year}{2010}).
	
	\bibitem[{\citenamefont{{Le Sueur} et~al.}(2010)\citenamefont{{Le Sueur},
	  Altimiras, Gennser, Cavanna, Mailly, and Pierre}}]{LeSueur.2010}
	\bibinfo{author}{\bibfnamefont{H.}~\bibnamefont{{Le Sueur}}},
	  \bibinfo{author}{\bibfnamefont{C.}~\bibnamefont{Altimiras}},
	  \bibinfo{author}{\bibfnamefont{U.}~\bibnamefont{Gennser}},
	  \bibinfo{author}{\bibfnamefont{A.}~\bibnamefont{Cavanna}},
	  \bibinfo{author}{\bibfnamefont{D.}~\bibnamefont{Mailly}}, \bibnamefont{and}
	  \bibinfo{author}{\bibfnamefont{F.}~\bibnamefont{Pierre}},
	  \bibinfo{journal}{{Physical Review Letters}} \textbf{\bibinfo{volume}{105}},
	  \bibinfo{pages}{056803} (\bibinfo{year}{2010}).
	
	\bibitem[{\citenamefont{Altimiras et~al.}(2010)\citenamefont{Altimiras, {Le
	  Sueur}, Gennser, Cavanna, Mailly, and Pierre}}]{Altimiras.2010}
	\bibinfo{author}{\bibfnamefont{C.}~\bibnamefont{Altimiras}},
	  \bibinfo{author}{\bibfnamefont{H.}~\bibnamefont{{Le Sueur}}},
	  \bibinfo{author}{\bibfnamefont{U.}~\bibnamefont{Gennser}},
	  \bibinfo{author}{\bibfnamefont{A.}~\bibnamefont{Cavanna}},
	  \bibinfo{author}{\bibfnamefont{D.}~\bibnamefont{Mailly}}, \bibnamefont{and}
	  \bibinfo{author}{\bibfnamefont{F.}~\bibnamefont{Pierre}},
	  \bibinfo{journal}{{Nature Physics}} \textbf{\bibinfo{volume}{6}},
	  \bibinfo{pages}{34} (\bibinfo{year}{2010}).
	
	\bibitem[{\citenamefont{Ferraro et~al.}(2014)\citenamefont{Ferraro, Roussel,
	  Cabart, Thibierge, F{\`e}ve, Grenier, and Degiovanni}}]{Ferraro.2014}
	\bibinfo{author}{\bibfnamefont{D.}~\bibnamefont{Ferraro}},
	  \bibinfo{author}{\bibfnamefont{B.}~\bibnamefont{Roussel}},
	  \bibinfo{author}{\bibfnamefont{C.}~\bibnamefont{Cabart}},
	  \bibinfo{author}{\bibfnamefont{E.}~\bibnamefont{Thibierge}},
	  \bibinfo{author}{\bibfnamefont{G.}~\bibnamefont{F{\`e}ve}},
	  \bibinfo{author}{\bibfnamefont{C.}~\bibnamefont{Grenier}}, \bibnamefont{and}
	  \bibinfo{author}{\bibfnamefont{P.}~\bibnamefont{Degiovanni}},
	  \bibinfo{journal}{{Physical Review Letters}} \textbf{\bibinfo{volume}{113}},
	  \bibinfo{pages}{166403} (\bibinfo{year}{2014}).
	
	\bibitem[{\citenamefont{Acciai et~al.}(2017)\citenamefont{Acciai, Calzona,
	  Dolcetto, Schmidt, and Sassetti}}]{Acciai.2017}
	\bibinfo{author}{\bibfnamefont{M.}~\bibnamefont{Acciai}},
	  \bibinfo{author}{\bibfnamefont{A.}~\bibnamefont{Calzona}},
	  \bibinfo{author}{\bibfnamefont{G.}~\bibnamefont{Dolcetto}},
	  \bibinfo{author}{\bibfnamefont{T.~L.} \bibnamefont{Schmidt}},
	  \bibnamefont{and} \bibinfo{author}{\bibfnamefont{M.}~\bibnamefont{Sassetti}},
	  \bibinfo{journal}{{Physical Review B}} \textbf{\bibinfo{volume}{96}},
	  \bibinfo{pages}{075144} (\bibinfo{year}{2017}).
	
	\bibitem[{\citenamefont{Cabart et~al.}(2018)\citenamefont{Cabart, Roussel,
	  F{\`e}ve, and Degiovanni}}]{Cabart.2018}
	\bibinfo{author}{\bibfnamefont{C.}~\bibnamefont{Cabart}},
	  \bibinfo{author}{\bibfnamefont{B.}~\bibnamefont{Roussel}},
	  \bibinfo{author}{\bibfnamefont{G.}~\bibnamefont{F{\`e}ve}}, \bibnamefont{and}
	  \bibinfo{author}{\bibfnamefont{P.}~\bibnamefont{Degiovanni}},
	  \bibinfo{journal}{{Physical Review B}} \textbf{\bibinfo{volume}{98}},
	  \bibinfo{pages}{155302} (\bibinfo{year}{2018}).
	
	\bibitem[{\citenamefont{Degiovanni et~al.}(2010)\citenamefont{Degiovanni,
	  Grenier, F{\`e}ve, Altimiras, {Le Sueur}, and Pierre}}]{Degiovanni.2010}
	\bibinfo{author}{\bibfnamefont{P.}~\bibnamefont{Degiovanni}},
	  \bibinfo{author}{\bibfnamefont{C.}~\bibnamefont{Grenier}},
	  \bibinfo{author}{\bibfnamefont{G.}~\bibnamefont{F{\`e}ve}},
	  \bibinfo{author}{\bibfnamefont{C.}~\bibnamefont{Altimiras}},
	  \bibinfo{author}{\bibfnamefont{H.}~\bibnamefont{{Le Sueur}}},
	  \bibnamefont{and} \bibinfo{author}{\bibfnamefont{F.}~\bibnamefont{Pierre}},
	  \bibinfo{journal}{{Physical Review B}} \textbf{\bibinfo{volume}{81}},
	  \bibinfo{pages}{121302} (\bibinfo{year}{2010}).
	
	\bibitem[{\citenamefont{Lunde et~al.}(2010)\citenamefont{Lunde, Nigg, and
	  Buttiker}}]{Lunde.2010}
	\bibinfo{author}{\bibfnamefont{A.~M.} \bibnamefont{Lunde}},
	  \bibinfo{author}{\bibfnamefont{S.~E.} \bibnamefont{Nigg}}, \bibnamefont{and}
	  \bibinfo{author}{\bibfnamefont{M.}~\bibnamefont{Buttiker}},
	  \bibinfo{journal}{{Physical Review B}} \textbf{\bibinfo{volume}{81}},
	  \bibinfo{pages}{041311} (\bibinfo{year}{2010}).
	
	\bibitem[{\citenamefont{Kovrizhin and Chalker}(2011)}]{Kovrizhin.2011}
	\bibinfo{author}{\bibfnamefont{D.~L.} \bibnamefont{Kovrizhin}}
	  \bibnamefont{and} \bibinfo{author}{\bibfnamefont{J.~T.}
	  \bibnamefont{Chalker}}, \bibinfo{journal}{{Physical Review B}}
	  \textbf{\bibinfo{volume}{84}}, \bibinfo{pages}{085105}
	  (\bibinfo{year}{2011}).
	
	\bibitem[{\citenamefont{Levkivskyi and Sukhorukov}(2012)}]{Levkivskyi.2012}
	\bibinfo{author}{\bibfnamefont{I.~P.} \bibnamefont{Levkivskyi}}
	  \bibnamefont{and} \bibinfo{author}{\bibfnamefont{E.~V.}
	  \bibnamefont{Sukhorukov}}, \bibinfo{journal}{{Physical Review B}}
	  \textbf{\bibinfo{volume}{85}}, \bibinfo{pages}{075309}
	  (\bibinfo{year}{2012}).
	
	\bibitem[{\citenamefont{Slobodeniuk et~al.}(2016)\citenamefont{Slobodeniuk,
	  Idrisov, and Sukhorukov}}]{Slobodeniuk.2016}
	\bibinfo{author}{\bibfnamefont{A.~O.} \bibnamefont{Slobodeniuk}},
	  \bibinfo{author}{\bibfnamefont{E.~G.} \bibnamefont{Idrisov}},
	  \bibnamefont{and} \bibinfo{author}{\bibfnamefont{E.~V.}
	  \bibnamefont{Sukhorukov}}, \bibinfo{journal}{{Physical Review B}}
	  \textbf{\bibinfo{volume}{93}}, \bibinfo{pages}{035421}
	  (\bibinfo{year}{2016}).
	
	\bibitem[{\citenamefont{Duprez et~al.}(2019)\citenamefont{Duprez, Sivre,
	  Anthore, Aassime, Cavanna, Ouerghi, Gennser, and Pierre}}]{Duprez.2019}
	\bibinfo{author}{\bibfnamefont{H.}~\bibnamefont{Duprez}},
	  \bibinfo{author}{\bibfnamefont{E.}~\bibnamefont{Sivre}},
	  \bibinfo{author}{\bibfnamefont{A.}~\bibnamefont{Anthore}},
	  \bibinfo{author}{\bibfnamefont{A.}~\bibnamefont{Aassime}},
	  \bibinfo{author}{\bibfnamefont{A.}~\bibnamefont{Cavanna}},
	  \bibinfo{author}{\bibfnamefont{A.}~\bibnamefont{Ouerghi}},
	  \bibinfo{author}{\bibfnamefont{U.}~\bibnamefont{Gennser}}, \bibnamefont{and}
	  \bibinfo{author}{\bibfnamefont{F.}~\bibnamefont{Pierre}},
	  \bibinfo{journal}{{Physical Review X}} \textbf{\bibinfo{volume}{9}},
	  \bibinfo{pages}{021030} (\bibinfo{year}{2019}).
	
	\bibitem[{\citenamefont{Inoue et~al.}(2014)\citenamefont{Inoue, Grivnin, Ofek,
	  Neder, Heiblum, Umansky, and Mahalu}}]{Inoue.2014}
	\bibinfo{author}{\bibfnamefont{H.}~\bibnamefont{Inoue}},
	  \bibinfo{author}{\bibfnamefont{A.}~\bibnamefont{Grivnin}},
	  \bibinfo{author}{\bibfnamefont{N.}~\bibnamefont{Ofek}},
	  \bibinfo{author}{\bibfnamefont{I.}~\bibnamefont{Neder}},
	  \bibinfo{author}{\bibfnamefont{M.}~\bibnamefont{Heiblum}},
	  \bibinfo{author}{\bibfnamefont{V.}~\bibnamefont{Umansky}}, \bibnamefont{and}
	  \bibinfo{author}{\bibfnamefont{D.}~\bibnamefont{Mahalu}},
	  \bibinfo{journal}{{Physical Review Letters}} \textbf{\bibinfo{volume}{112}},
	  \bibinfo{pages}{166801} (\bibinfo{year}{2014}).
	
	\bibitem[{\citenamefont{Itoh et~al.}(2018)\citenamefont{Itoh, Nakazawa, Ota,
	  Hashisaka, Muraki, and Fujisawa}}]{Itoh.2018}
	\bibinfo{author}{\bibfnamefont{K.}~\bibnamefont{Itoh}},
	  \bibinfo{author}{\bibfnamefont{R.}~\bibnamefont{Nakazawa}},
	  \bibinfo{author}{\bibfnamefont{T.}~\bibnamefont{Ota}},
	  \bibinfo{author}{\bibfnamefont{M.}~\bibnamefont{Hashisaka}},
	  \bibinfo{author}{\bibfnamefont{K.}~\bibnamefont{Muraki}}, \bibnamefont{and}
	  \bibinfo{author}{\bibfnamefont{T.}~\bibnamefont{Fujisawa}},
	  \bibinfo{journal}{{Physical Review Letters}} \textbf{\bibinfo{volume}{120}},
	  \bibinfo{pages}{197701} (\bibinfo{year}{2018}).
	
	\bibitem[{\citenamefont{Imambekov and Glazman}(2009)}]{Imambekov.2009}
	\bibinfo{author}{\bibfnamefont{A.}~\bibnamefont{Imambekov}} \bibnamefont{and}
	  \bibinfo{author}{\bibfnamefont{L.~I.} \bibnamefont{Glazman}},
	  \bibinfo{journal}{{Science (New York, N.Y.)}} \textbf{\bibinfo{volume}{323}},
	  \bibinfo{pages}{228} (\bibinfo{year}{2009}).
	
	\bibitem[{\citenamefont{Imambekov et~al.}(2012)\citenamefont{Imambekov,
	  Schmidt, and Glazman}}]{Imambekov.2012}
	\bibinfo{author}{\bibfnamefont{A.}~\bibnamefont{Imambekov}},
	  \bibinfo{author}{\bibfnamefont{T.~L.} \bibnamefont{Schmidt}},
	  \bibnamefont{and} \bibinfo{author}{\bibfnamefont{L.~I.}
	  \bibnamefont{Glazman}}, \bibinfo{journal}{{Reviews of Modern Physics}}
	  \textbf{\bibinfo{volume}{84}}, \bibinfo{pages}{1253} (\bibinfo{year}{2012}).
	
	\bibitem[{\citenamefont{Karzig et~al.}(2010)\citenamefont{Karzig, Glazman, and
	  von Oppen}}]{Karzig.2010}
	\bibinfo{author}{\bibfnamefont{T.}~\bibnamefont{Karzig}},
	  \bibinfo{author}{\bibfnamefont{L.~I.} \bibnamefont{Glazman}},
	  \bibnamefont{and} \bibinfo{author}{\bibfnamefont{F.}~\bibnamefont{von
	  Oppen}}, \bibinfo{journal}{{Physical Review Letters}}
	  \textbf{\bibinfo{volume}{105}}, \bibinfo{pages}{226407}
	  (\bibinfo{year}{2010}).
	
	\bibitem[{\citenamefont{Degiovanni et~al.}(2009)\citenamefont{Degiovanni,
	  Grenier, and F{\`e}ve}}]{Degiovanni.2009}
	\bibinfo{author}{\bibfnamefont{P.}~\bibnamefont{Degiovanni}},
	  \bibinfo{author}{\bibfnamefont{C.}~\bibnamefont{Grenier}}, \bibnamefont{and}
	  \bibinfo{author}{\bibfnamefont{G.}~\bibnamefont{F{\`e}ve}},
	  \bibinfo{journal}{{Physical Review B}} \textbf{\bibinfo{volume}{80}},
	  \bibinfo{pages}{241307(R)} (\bibinfo{year}{2009}).
	
	\bibitem[{\citenamefont{Chalker et~al.}(2007)\citenamefont{Chalker, Gefen, and
	  Veillette}}]{Chalker.2007}
	\bibinfo{author}{\bibfnamefont{J.~T.} \bibnamefont{Chalker}},
	  \bibinfo{author}{\bibfnamefont{Y.}~\bibnamefont{Gefen}}, \bibnamefont{and}
	  \bibinfo{author}{\bibfnamefont{M.~Y.} \bibnamefont{Veillette}},
	  \bibinfo{journal}{{Physical Review B}} \textbf{\bibinfo{volume}{76}},
	  \bibinfo{pages}{085320} (\bibinfo{year}{2007}).
	
	\bibitem[{\citenamefont{Neuenhahn and Marquardt}(2008)}]{Neuenhahn.2008}
	\bibinfo{author}{\bibfnamefont{C.}~\bibnamefont{Neuenhahn}} \bibnamefont{and}
	  \bibinfo{author}{\bibfnamefont{F.}~\bibnamefont{Marquardt}},
	  \bibinfo{journal}{{New Journal of Physics}} \textbf{\bibinfo{volume}{10}},
	  \bibinfo{pages}{115018} (\bibinfo{year}{2008}).
	
	\bibitem[{\citenamefont{Neuenhahn and Marquardt}(2009)}]{Neuenhahn.2009}
	\bibinfo{author}{\bibfnamefont{C.}~\bibnamefont{Neuenhahn}} \bibnamefont{and}
	  \bibinfo{author}{\bibfnamefont{F.}~\bibnamefont{Marquardt}},
	  \bibinfo{journal}{{Physical Review Letters}} \textbf{\bibinfo{volume}{102}},
	  \bibinfo{pages}{046806} (\bibinfo{year}{2009}).
	
	\bibitem[{\citenamefont{Heyl et~al.}(2010)\citenamefont{Heyl, Kehrein, Marquardt, and Neuenhahn}}]{Heyl.2010}
	\bibinfo{author}{\bibfnamefont{M.}~\bibnamefont{Heyl}},
	  \bibinfo{author}{\bibfnamefont{S.}~\bibnamefont{Kehrein}},
	  \bibinfo{author}{\bibfnamefont{F.}~\bibnamefont{Marquardt}}, \bibnamefont{and}
	  \bibinfo{author}{\bibfnamefont{C}~\bibnamefont{Neuenhahn}},
	  \bibinfo{journal}{{Physical Review B}} \textbf{\bibinfo{volume}{82}},
	  \bibinfo{pages}{033409} (\bibinfo{year}{2010}).

	\bibitem[{\citenamefont{Karzig}()}]{Karzig.2012}
	\bibinfo{author}{\bibfnamefont{T.}~\bibnamefont{Karzig}},
	  \emph{\bibinfo{title}{{Low dimensional electron systems out of
	  equilibrium}}}, \bibinfo{note}{(Dissertation, Freie Universit{\"a}t Berlin,
	  Berlin, 2012)}.
	
	\bibitem[{\citenamefont{Fischer et~al.}(2019)\citenamefont{Fischer, Park, Meir,
	  and Gefen}}]{Fischer.2019}
	\bibinfo{author}{\bibfnamefont{S.~G.} \bibnamefont{Fischer}},
	  \bibinfo{author}{\bibfnamefont{J.}~\bibnamefont{Park}},
	  \bibinfo{author}{\bibfnamefont{Y.}~\bibnamefont{Meir}}, \bibnamefont{and}
	  \bibinfo{author}{\bibfnamefont{Y.}~\bibnamefont{Gefen}},
	  \bibinfo{journal}{{Physical Review B}} \textbf{\bibinfo{volume}{100}},
	  \bibinfo{pages}{195411} (\bibinfo{year}{2019}).
	
	\bibitem[{\citenamefont{Kane and Fisher}(2003)}]{Kane.2003}
	\bibinfo{author}{\bibfnamefont{C.~L.} \bibnamefont{Kane}} \bibnamefont{and}
	  \bibinfo{author}{\bibfnamefont{M.~P.~A.} \bibnamefont{Fisher}},
	  \bibinfo{journal}{{Physical Review B}} \textbf{\bibinfo{volume}{67}},
	  \bibinfo{pages}{045307} (\bibinfo{year}{2003}).
	
	\bibitem[{\citenamefont{Takei et~al.}(2010)\citenamefont{Takei, Milletar{\`i},
	  and Rosenow}}]{Takei.2010}
	\bibinfo{author}{\bibfnamefont{S.}~\bibnamefont{Takei}},
	  \bibinfo{author}{\bibfnamefont{M.}~\bibnamefont{Milletar{\`i}}},
	  \bibnamefont{and} \bibinfo{author}{\bibfnamefont{B.}~\bibnamefont{Rosenow}},
	  \bibinfo{journal}{{Physical Review B}} \textbf{\bibinfo{volume}{82}},
	  \bibinfo{pages}{041306(R)} (\bibinfo{year}{2010}).
	
	\bibitem[{\citenamefont{Han et~al.}(2016)\citenamefont{Han, Park, Gefen, and
	  Sim}}]{Han.2016}
	\bibinfo{author}{\bibfnamefont{C.}~\bibnamefont{Han}},
	  \bibinfo{author}{\bibfnamefont{J.}~\bibnamefont{Park}},
	  \bibinfo{author}{\bibfnamefont{Y.}~\bibnamefont{Gefen}}, \bibnamefont{and}
	  \bibinfo{author}{\bibfnamefont{H.-S.} \bibnamefont{Sim}},
	  \bibinfo{journal}{{Nature communications}} \textbf{\bibinfo{volume}{7}},
	  \bibinfo{pages}{11131} (\bibinfo{year}{2016}).
	
	\end{thebibliography}
\end{document}